\documentclass[%
 reprint,
showpacs,
preprintnumbers,
 amsmath,amssymb,
 aps,
pra,
]{revtex4-1}

\usepackage{graphicx}% Include figure files
\usepackage{dcolumn}% Align table columns on decimal point
\usepackage{bm}% bold math
\usepackage{amsmath}  % needed for \tfrac, \bmatrix, etc.
\usepackage{amsfonts} % needed for bold Greek, Fraktur, and blackboard bold
\usepackage{amsmath,amssymb,epsfig}
\usepackage{graphicx}
\usepackage[vcentermath,stdtext]{youngtab}
\usepackage{epstopdf}
\begin{document}

\preprint{APS/123-QED}

\title{Realization of phase modulation using a spatial light modulator in the quantum Talbot configuration}

\author{S. Deachapunya}
\email{sarayut@buu.ac.th}
\affiliation{Department of Physics, Faculty of Science, Burapha University, ChonBuri Province, 20131, Thailand}

\author{S. Srisuphaphon}
\affiliation{Department of Physics, Faculty of Science, Burapha University, ChonBuri Province, 20131, Thailand}

\author{P. Panthong}
\affiliation{Department of Physics, Faculty of Science, Kasetsart University, Bangkok Province, 10900, Thailand}

\author{T. Photia}
\affiliation{Department of Physics, Faculty of Science, Kasetsart University, Bangkok Province, 10900, Thailand}

\author{K. Boonkham}
\affiliation{Department of Physics, Faculty of Science, Kasetsart University, Bangkok Province, 10900, Thailand}

\author{S. Chiangga}
\affiliation{Department of Physics, Faculty of Science, Kasetsart University, Bangkok Province, 10900, Thailand}

%\date{today}

\begin{abstract}
We demonstrate the quantum Talbot effect using pairs of single photons produced by parametric down conversion. In contrast to the previous works, we use a programmable spatial light modulator to behave as a diffraction grating. Thus, the investigation of the Talbot diffraction patterns under the variation of grating structure can be easily performed. The influence of spectral bandwidth of the down-converted photons on the diffraction pattern is also investigated. A theoretical model based on the wave nature of photons is presented to explain the Talbot diffraction pattern under varying conditions. The measured diffraction patterns are in good agreement with the theoretical prediction. We are convinced that our study improves the understanding of the quantum Talbot effect.
\end{abstract}

\maketitle

\section{Introduction}

The Talbot effect, which was introduced in 1836 by Henry Fox Talbot~\cite{Talbot1836}, is a near-field diffraction effect. The Talbot diffraction patterns appear when a coherent light illuminates a diffraction grating. The grating self-image is formed behind the grating at multiples of specific distance, so-called the Talbot length. The experiments of near-field or Talbot effects have been implemented, including light~\cite{Case2009}, atom~\cite{Chapman1995}, molecule~\cite{Brezger2002}, surface plasmon~\cite{Dennis2007}, and non-linear optics~\cite{Bortolozzo2006}.

The diffraction of entangled photons has been numerously studied in the context of the superposition of two Bessel beams~\cite{Vidal2008}, quantum lithography~\cite{D'Angelo2001,Luo2009}, quantum ghost imaging~\cite{Strekalov1995}, and with the experimental Talbot effect~\cite{Song2011,Jin2012,Wen2013}. The first experiment of quantum Talbot effect~\cite{Song2011} used both single photons and biphoton pairs which were generated by spontaneous parametric down conversion (SPDC). The experimental configuration is similar to previous quantum ghost imaging setup~\cite{Strekalov1995}. The signal and idler photons from SPDC were split into two separate optical paths. The grating was placed in the signal path before a detector. The Talbot self-imaging of the grating was observed by scanning either the signal or the idler detector. In contrast to those studies, the quantum decoherence experiment has also been performed with single photons~\cite{D'Auria2011}.

A liquid crystal spatial light modulator (SLM) is a versatile device in a wide range of applications in optics, including the creation of holograms~\cite{Ahrenberg2006}, polarization control~\cite{Moreno2012}, and the study of optical vortices~\cite{Panthong2016}. The SLM can control both amplitude and phase of the incident light.

In this paper we report on the experimental and theoretical investigation of the quantum Talbot effect using pairs of single photons~\cite{Kwiat1999}. In contrast to the previous demonstrations~\cite{Song2011}, the SLM in our experimental setup is programmed to act as the diffraction grating. Our technique permits the real-time variation of the period and the opening fraction of grating. We further investigate the effect of spectral bandwidth down-converted photons on the diffraction patterns. In order to explain the Talbot diffraction pattern under varying conditions, the theoretical model based on paraxial approximation is presented in the next section. We subsequently present our experimental setup, and compare the results to our theoretical findings.

\section{Theory and method}

\begin{figure}[htb]
\begin{center}
\centering\includegraphics[width=1\columnwidth]{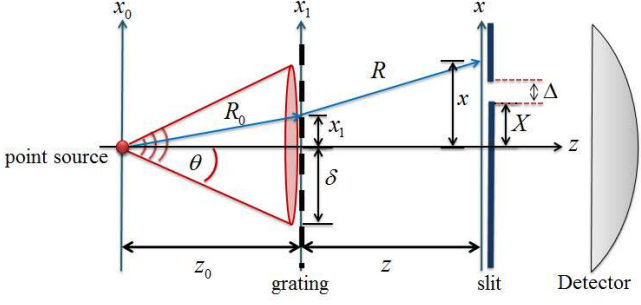}
\end{center}
\caption{A spherical wave of signal photon beam with the divergence angle about, $\theta \approx \delta/z_0$=0.5 mrad, diffracts through the grating which has the periodic modulation along the $x_1$-axis to a single slit and a detector is placed behind this slit for photon detection.}
\label{Fig1}
\end{figure}

In this section, we present a theoretical consideration of the Talbot effect for photon point source. Let a spherical wave from a point source encounters a grating at distance $z_0$ as shown in Figure~\ref{Fig1}.
Considering  the propagation in $xz$ plane, the field distribution with the leading order approximation $R_0\simeq z_0+x_1^2/2z_0$, can be obtained as
\begin{eqnarray} \label{initail wave1}
\psi_{1,-}(x_1,z_0)= {\rm exp} \{-i k \big(z_0 +\frac{ x_1^2}{2 z_0}\big)\},
\end{eqnarray}
where $k=2 \pi / \lambda $ is the magnitude of the wave vector with wavelength $\lambda$. The field is then incident onto the grating at $z_0$. For the diffraction grating, which has the periodic modulation along the $x_1$-axis with the period $d$, the wave is transformed to be
\begin{eqnarray} \label{initial G wave}
\psi_{1,+}(x_1,z_0)=\sum_{n}A_n {\rm exp}\{i n k_d x_1\}\psi_{1,-}(x_1,z_0),
\end{eqnarray}
where $n=0,\pm1,\pm2,...$, and $k_d=2 \pi /d $. The factor $A_n=\sin(n \pi f) / n \pi$  is the components of the Fourier decomposition of the periodic for the grating  with an opening fraction $f$~\cite{Case2009}. Subsequently, we apply the Huygens-Fresnel integral to find the wave function behind the grating at distance $z$ with the transverse axis $x$. The field distribution $\psi(x,z)$ is given by
\begin{eqnarray} \label{wave1}
\psi(x,z)&=&\sqrt{\frac{i}{\lambda }} \int_{-\infty}^{\infty}dx_1 \frac{{\rm exp}\{ - ik(z+\frac{(x-x_1)^2}{2z})\}}{z}
\nonumber
\\ && \times\psi_{1,+}(x_1,z_0).
\end{eqnarray}
According to the field $\psi_{1,+}(x_1,z_0)$ in Eq. (\ref{initial G wave}), the integral can be done analytically and the result is,
\begin{eqnarray} \label{wave2}
\psi(x,z)&=&\sqrt{\frac{ Z}{z^2}} {\rm exp}  \{-i k \big(z_0+z +\frac{ x_1^2}{2 z}\big)\}
\nonumber
\\ && \times\sum_{n}A_n{\rm exp}\{ \frac{i Z}{2k} \big(n k_d + \frac{kx}{z}\big)^2 \} ,
\end{eqnarray}
where the distance $Z=z z_0/(z+z_0)$. Omitting the pattern independent factor $\sqrt{Z/z^2}$, we obtain the intensity distribution $\psi^*\psi$ behind the grating in the form
\begin{eqnarray} \label{Intensity}
I(x,z)&=&\sum_{n,m}A_nA_m {\rm exp}\{i \Big((n-m) (\frac{k_d}{1+z/z_0}) x
\nonumber
\\ && + \frac{(n^2-m^2)\pi \lambda Z}{d^2}\Big) \}.
\end{eqnarray}
The obtained intensity can be reduced to the case of a plane wave. By taking $z_0\rightarrow\infty$, we therefore obtain the effective distance becomes $Z=z$. The interference pattern, which is similar to the period of the grating itself can be found when the longitudinal distance $z$ equals to an integer number of the Talbot length $L_T=d^2/\lambda$.

In order to compare the theoretical model to the experimental data, the effects due to quantum behaviour of single photons have to be considered.  Namely, the intensity pattern is corresponding to the probability distribution of the detecting photons along the $x$-axis. Therefore, the obtained intensity (Eq.(\ref{Intensity})) can be considered as the number of incident photons. One can evaluate the count rate of photons propagated through a single slit by integrating the intensity over the slit interval, $\int_{X}^{X+\Delta}\psi^*\psi dx$, where $X$ denotes the position of slit with the slit width, $\Delta$ as shown in Figure~\ref{Fig1}. With the intensity distribution in Eq.(\ref{Intensity}), the count rate in this scheme can be obtained as
\begin{eqnarray} \label{CR}
P_{\lambda}(X)&\propto&\sum_{n,m}A_nA_m \frac{\sin((n-m)(\frac{k_d}{1+z/z_0}) (\Delta /2))}{(n-m)(\frac{k_d}{1+z/z_0}) }
\nonumber
\\ && \times {\rm exp}\{i \Big((n-m) (\frac{k_d}{1+z/z_0}) (X+\frac{\Delta}{2} )
\nonumber
\\ && +\frac{(n^2-m^2)\pi \lambda Z}{d^2}\Big) \}.
\end{eqnarray}
A single photon detector is placed behind the slit (Figure~\ref{Fig1}). The interference pattern can then be revealed by scanning transversely the grating over the slit or vice versa~\cite{Deachapunya2014b}.

In addition, the effect of the wavelength distribution has to be involved. For a Gaussian distribution of the wavelength around $\lambda_0 $, the total count rate is given by
\begin{eqnarray} \label{probability}
P(X)=\sum^\infty_{\lambda=0}e^{-\frac{(\lambda-\lambda_0)^2}{\beta^2}}P_{\lambda}(X),
\end{eqnarray}
where $\beta$ represents the radius of the Gaussian distribution.

\section{Experiment}

\begin{figure}[htb]
\centering\includegraphics[width=1\columnwidth]{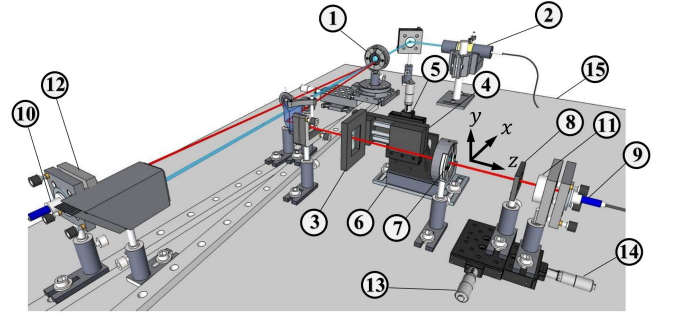}
\caption{(Color online) Our quantum Talbot setup with photon pairs. (1) BBO; (2) 50 mW laser diode ($\lambda$=405 nm); (3) SLM; (4) first translation stage (MTS50/M-Z8E, Thorlabs); (5) second translation stage (PT1/M, Thorlabs); (6) third translation stage (Z812B, Thorlabs); (7) polarizer; (8) single slit; (9) optical fiber for signal photon APD; (10) optical fiber for idler photon APD; (11) signal photon filter; (12) idler photon filter; (13) single slit stage (PT1/M); (14) stage for signal photon APD; (15) optical table. The photon pairs of the wavelength 810 nm are produced from spontaneous parametric down conversion. A SLM is used in the grating mode: the details are described in the text.}
\label{Fig2}
\end{figure}

We set the quantum Talbot experiment with the single photon source. Our single photon pairs are produced from spontaneous parametric down conversion with type I configuration. A beta barium borate (BBO) is pumped by a 50 mW diode laser (LQA405-50E, Newport) with the center wavelength of 405 nm, diameter of about 1.5 mm. The angle between the signal and idler photons of 6 degrees is selected in order to obtain the photon pairs of the wavelength 810 nm (Figure~\ref{Fig2}). Each of photon pairs has the divergence angle ($\theta$) of about 0.5 mrad. The signal photons encounter the SLM (LC2012, Holoeye Photonics AG with resolution of 1024$\times$768 pixels, the pixel size of 36 $\mu$m), using as a diffraction grating, which can be varied both the period and the opening fraction. The azimuthal phase of the SLM varies from 0 to around 1$\pi$ with 8 bit (256 gray levels) level at 810 nm of the photon source. An unmodulated phase of photons can be blocked by a polarizer. The SLM application software is used to the grating mode. The grating mode of the period, $d$=360 $\mu$m, and the opening fractions (the ratio between the grating window and the grating period) of $f$=0.1-0.5 is set for the experiments. The typical Talbot length is then calculated to be $L_{T}$=160 mm. The SLM is mounted on three translation stages. The first one (MTS50/M-Z8E, resolution 1.6 $\mu$m, Thorlabs) can move along the longitudinal direction in order to set the distance $z$. The second one (PT1/M, Thorlabs) can move the SLM in the y axis for adjusting the center of the grating. The third one (Z812B, Thorlabs) is used to explore the interference patterns automatically by scanning the SLM along the transverse x axis with the step of 12 $\mu$m. The coincidence count is recorded for each transverse (x) scan. Behind the SLM grating, a single slit with the width of about 115 $\mu$m is placed at the distance of $z$=160 mm, and subsequently a fiber-coupled avalanche photodiode module (SPCM-AQ4C, PerkinElmer) is used as a single photon detection (Figure~\ref{Fig2}). The single slit can be eliminated by a manual stage (PT1/M, Thorlabs) when the photon alignment is needed, as well as the APD can be roughly adjusted the longitudinal distance by another stage (PT1/M, Thorlabs). The idler photon is employed as a trigger to ensure the signal photon from noises. The time window of 25 ns is applied for coincidence counts. A long pass filter with FWHM of 50 nm is placed at front of both photon pairs. The broad filters are used for producing the broad wavelength photon source.

\section{Results and discussions}

The effect of a point source dominates the phase modulation of the photon wave function and the interference patterns are therefore expanded and the effective Talbot length has been extended to 174 mm, which is longer than the typical Talbot length of $L_{T}$=160 mm.

\begin{figure*}[htb]
\centering
\includegraphics[width=11cm]{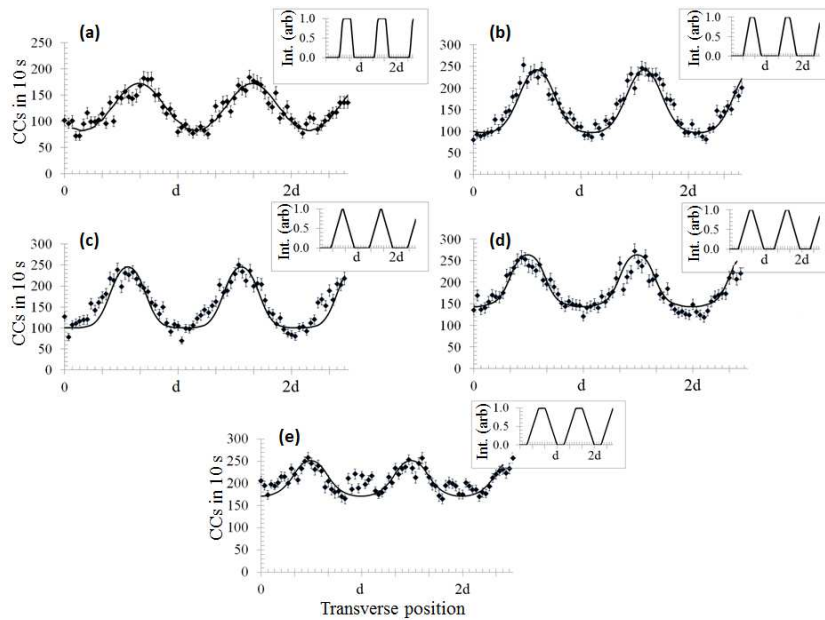}
\caption{The experimental interference patterns at $z$=160 mm, $d$=360 $\mu$m, $\theta$=0.5 mrad, $\lambda$=810 nm, and FWHM = 50nm (detail see text): $f$=0.1 (a), $f$=0.2 (b), $f$=0.3 (c), $f$=0.4 (d), $f$=0.5 (e). The solid lines represent the theory done with Eq.(\ref{probability}). The coincidence counts (CCs) of photon pairs were accumulated in 10 s. The transverse position (x) represents in the unit of grating period (d). The error bars on the y-axis represent the photon shot noise. The error bars on the x-axis are small. The insets show the simulation of the normalized intensity pattern with the condition of using a plane-parallel and monochromatic beam.}
\label{Fig3}
\end{figure*}

Figure~\ref{Fig3} presents the interference patterns of the photons from the quantum Talbot effect at $z$=160 mm. The period of the grating, $d$=360 $\mu$m, the center wavelength, $\lambda$=810 nm, FWHM of 50 nm, and the single slit with the width of about 115 $\mu$m were used. Our signal photon beam has the divergence angle ($\theta$) of about 0.5 mrad. In details, Figure~\ref{Fig3} (a) is the pattern with the opening fraction, $f$=0.1. The fringe width is about 47\%, which is larger than a regular width of 10\%. This also happens throughout the other opening fractions, $f$=0.2-0.5 (Figure~\ref{Fig3} (b)-(e)). A controlled visibility can be obviously observed with different opening fractions. We check the results with the simulations done with Eq.(\ref{probability}), and present by the solid lines in Figure~\ref{Fig3}. They are in good agreement with each other. We also show the simulations with the same conditions as before but using a plane-parallel ($\theta$=0 rad) and monochromatic (FWHM = 0 nm) beam. They are presented as the insets in Figure~\ref{Fig3}. The width of the interference patterns becomes larger with the higher opening fractions. This cannot be seen with our recent experiments because the reason of broad beam and wavelength distribution. In our study here, we include various effects, i.e. wavelength distribution, beam divergence, opening fraction, and single slit at the detector. All these effects influence on the phase modulation of the wave function as described in our theory and we can observe in our experimental fringe patterns.

\section{Conclusion}

The Talbot effect has been studied with a single photon source. A series of experiments was conducted for different opening fractions of the grating with the help of SLM. The use of SLM in the single photon Talbot experiment has merits. The phase modulation can be controlled and observed. Our results show that the measured diffraction widths become wider as the filter bandwidths increase. In addition, the visibility of the interference patterns can be controlled with the opening fractions of the grating. The calculations are used to verify the experiments and fit to the data quite well. We hope that our study here is useful and provides a better understanding for quantum experiments, as well as further implementation of quantum spectroscopy or small signal spectroscopy such as molecule spectroscopy.

\begin{acknowledgements}
 S.D. acknowledges the support grant from the office of the higher education commission, the Thailand research fund (TRF), and Faculty of Science, Burapha University under contract number MRG5380264. We thank W. Temnuch for her help in a part of photon alignment.
\end{acknowledgements}

%\bibliographystyle{unsrt}
%\bibliography{/home/thiago/bibtex/articles,/home/thiago/bibtex/books}

\end{document}